\documentclass[graybox]{svmult}

\usepackage{mathptmx}       
\usepackage{helvet}         
\usepackage{courier}        
\usepackage{type1cm}        
\usepackage{makeidx}         
\usepackage{graphicx}        
\usepackage{multicol}        
\usepackage[bottom]{footmisc}

\usepackage{amsmath}
\usepackage{amssymb}

\newcommand{\vect}[1]{\vec{#1}}
\newcommand{\tensor}[1]{\mathbf{#1}}
\newcommand{\numflux}{\mathcal{F}}

\makeindex             

\begin{document}

\title*{Simulating Turbulence Using the Astrophysical Discontinuous Galerkin Code TENET}
\titlerunning{Simulating Turbulence Using the Astrophysical DG Code TENET}
\author{Andreas Bauer, Kevin Schaal, Volker Springel, Praveen Chandrashekar, R\"udiger Pakmor and Christian Klingenberg}
\authorrunning{A. Bauer et al.}
\institute{Andreas Bauer \textbullet~ R\"udiger Pakmor \at Heidelberg Institute for Theoretical Studies, Schloss-Wolfsbrunnenweg 35, 69118 Heidelberg, Germany,\\ \email{andreas.bauer@h-its.org; ruediger.pakmor@h-its.org}
\and Kevin Schaal \textbullet~ Volker Springel \at Heidelberg Institute for Theoretical Studies, Schloss-Wolfsbrunnenweg 35, 69118 Heidelberg, Germany,\\ Zentrum f\"ur Astronomie der Universit\"at Heidelberg,
  Astronomisches Recheninstitut, M\"{o}nchhofstr. 12-14, 69120
  Heidelberg, Germany,\\ \email{kevin.schaal@h-its.org; volker.springel@h-its.org}
\and
  Praveen Chandrashekar \at TIFR Centre for Applicable Mathematics, Bangalore-560065, India,\\ \email{praveen@tifrbng.res.in}
  \and Christian Klingenberg \at Institut f\"ur Mathematik, Universit\"at W\"urzburg,
  Emil-Fischer-Str. 30, 97074 W\"urzburg, Germany,\\ \email{klingenberg@mathematik.uni-wuerzburg.de}}
%
%
\maketitle

\abstract*{In astrophysics, the two main methods traditionally in use
  for solving the Euler equations of ideal fluid dynamics are smoothed
  particle hydrodynamics and finite volume discretization on a
  stationary mesh. However, the goal to efficiently make use of future
  exascale machines with their ever higher degree of parallel
  concurrency motivates the search for more efficient and more
  accurate techniques for computing hydrodynamics. Discontinuous
  Galerkin (DG) methods represent a promising class of methods in this
  regard, as they can be straightforwardly extended to arbitrarily
  high order while requiring only small stencils. Especially for
  applications involving comparatively smooth problems, higher-order
  approaches promise significant gains in computational speed for
  reaching a desired target accuracy. Here, we introduce our new
  astrophysical DG code {\small TENET} designed for applications in
  cosmology, and discuss our first results for 3D simulations of subsonic
  turbulence. We show that our new DG implementation provides accurate
  results for subsonic turbulence, at considerably reduced
  computational cost compared with traditional finite volume
  methods. In particular, we find that DG needs about 1.8 times fewer
  degrees of freedom to achieve the same accuracy and at the same time
  is more than 1.5 times faster, confirming its substantial promise
  for astrophysical applications.}

\abstract{In astrophysics, the two main methods traditionally in use
  for solving the Euler equations of ideal fluid dynamics are smoothed
  particle hydrodynamics and finite volume discretization on a
  stationary mesh. However, the goal to efficiently make use of future
  exascale machines with their ever higher degree of parallel
  concurrency motivates the search for more efficient and more
  accurate techniques for computing hydrodynamics. Discontinuous
  Galerkin (DG) methods represent a promising class of methods in this
  regard, as they can be straightforwardly extended to arbitrarily
  high order while requiring only small stencils. Especially for
  applications involving comparatively smooth problems, higher-order
  approaches promise significant gains in computational speed for
  reaching a desired target accuracy. Here, we introduce our new
  astrophysical DG code {\small TENET} designed for applications in
  cosmology, and discuss our first results for 3D simulations of subsonic
  turbulence. We show that our new DG implementation provides accurate
  results for subsonic turbulence, at considerably reduced
  computational cost compared with traditional finite volume
  methods. In particular, we find that DG needs about 1.8 times fewer
  degrees of freedom to achieve the same accuracy and at the same time
  is more than 1.5 times faster, confirming its substantial promise
  for astrophysical applications.}

\section{Introduction}

Turbulent flows are ubiquitous in astrophysical systems. For example,
supersonic turbulence in the interstellar medium is thought to play a
key role in regulating star formation
\cite{Klessen2000, MacLow2004}. In cosmic structure formation,
turbulence occurs in accretion flows onto halos and contributes to the
pressure support in clusters of galaxies \cite{Schuecker2004} and
helps in distributing and mixing heavy elements into the primordial
gas. Also, turbulence plays a crucial role in creating an effective
viscosity and mediating angular momentum transport in gaseous
accretion flows around supermassive black holes.

Numerical simulations of astrophysical turbulence require an accurate
treatment of the Euler equations. Traditionally, finite volume schemes
have been used in astrophysics for high accuracy simulations of
hydrodynamics. They are mostly based on simple linear data
reconstruction resulting in second-order accurate schemes.  In
principle, these finite volume schemes can also be extended to high
order, with the next higher order method using parabolic data
reconstruction, as implemented in piecewise parabolic schemes
\cite{Colella1984}. While a linear reconstruction needs only the
direct neighbours of each cell, a further layer is required for the
parabolic reconstruction. In general, with the increase of the order
of the finite volume scheme, the required stencil grows as
well. Especially in a parallelized code, this affects the scalability,
as the ghost region around the local domain has to grow as well for a
deeper stencil, resulting in larger data exchanges among different MPI
processes and higher memory overhead.

An interesting and still comparatively new alternative are so-called
discontinuous Galerkin (DG) methods. They rely on a representation of
the solution through an expansion into basis functions at the sub-cell
level, removing the reconstruction necessary in high-order finite
volume schemes.  Such DG methods were first introduced by
\cite{Reed1973}, and later extended to non-linear problems
\cite{Cockburn1989a, Cockburn1989b, Cockburn1990, Cockburn1991,
  Cockburn1998}. Successful applications have so far been mostly
reported for engineering problems \cite{Cockburn2011,
  Gallego-Valencia2014}, but they have very recently also been
considered for astrophysical problems \cite{Mocz2014, Zanotti2015}.
DG methods only need information about their direct neighbours,
independent of the order of the scheme. Furthermore, the computational
workload is not only spent on computing fluxes between cells, but has
a significant internal contribution from each cell as well. The latter
part is much easier to parallelize in a hybrid parallelization
code. Additionally, DG provides a systematic and transparent framework
to derive discretized equations up to an arbitrarily high
convergence order. These features make DG methods a compelling
approach for future exa-scale machines. Building higher order methods
with a classical finite volume approach is rather contrived in
comparison, which is an important factor in explaining why mostly
second and third order finite volume methods are used in practice.

As shown in \cite{Bauer2012}, subsonic turbulence can pose a hard
problem for some of the simulation methods used in computational
astrophysics.  Standard SPH in particular struggles to reproduce
results as accurate as finite volume codes, and a far higher
computational effort would be required to obtain an equally large
inertial range as obtained with a finite volume method, a situation
that has only been moderately improved by many enhancements proposed
for SPH in recent years \cite{Wadsley2008, Price2008, Hess2010,
  Read2010, Abel2011, Hopkins2013}. In this work, we explore instead
how well the DG methods implemented in our new astrophysical
simulation code {\small TENET} \cite{Schaal2015} perform for
simulations of subsonic turbulence.  In this problem, the
discontinuities between adjacent cells are expected to be small and
the sub-cell representation within a cell can reach high
accuracy. This makes subsonic turbulence a very interesting first
application of our new DG implementation.

In the following, we outline the equations and main ideas behind DG
and introduce our implementation. We will first describe how the
solution is represented using a set of basis functions. Then, we
explain how initial conditions can be derived and how they are evolved
forward in time. Next, we examine how well our newly developed DG
methods behave in simulating turbulent flows. In particular, we test
whether an improvement in accuracy and computational efficiency
compared with standard second-order finite volume methods is indeed
realized.

\section{Discontinuous Galerkin methods} 

Galerkin methods form a large class of methods for converting
  continuous differential equations into sets of discrete differential
  equations \cite{Galerkin1915}. Instead of describing the solution
with averaged quantities $\vect{q}$ within each cell, in DG the
solution is represented by an expansion into basis functions, which
are often chosen as polynomials of degree $k$.  This polynomial
representation is continuous inside a cell, but discontinuous across
cells, hence the name discontinuous Galerkin method.  Inside a cell
$K$, the state is described by a function
$\vect{q}^{K}(\vect{x},t)$. This function is only defined on the
volume of cell $K$. In the following, we will use $\vect{q}^K$ to
refer to the polynomial representation of the state inside cell $K$.

The polynomials of degree $k$ form a vector space, and the state
$\vect{q}^K$ within a cell can be represented using weights
$\vect{w}_l^K$, where $l$ denotes the component of the weight
vector. Each $\vect{w}_l$ contains an entry for each of the five
conserved hydrodynamic quantities. Using a set of suitable orthogonal
basis functions $\phi_l^K(\vect{x})$, the state in a cell can be
expressed as
\begin{align}
\vect{q}^K\left(\vect{x},t\right)=\sum\limits_{l=1}^{N(k)}\vect{w}_l^K(t)\phi_l^K(\vect{x}).
\end{align}
Note how the time and space dependence on the right hand side is split up
into two functions. This will provide the key ingredient for discretizing
the continuous partial differential equations into a set of coupled
ordinary differential equations.

The vector space of all polynomials up to degree $k$ has the dimension
$N(k)$. The $l$-th component of the vector can be obtained through a
projection of the state $\vect{q}$ onto the $l$-th basis function:
\begin{align} \vect{w}_l^K(t) = \frac{1}{|K|}\int_K\!\vect{q}(\vect{x},t)
\phi^K_l(\vect{x})\,\mathrm{d}V, \end{align} with $|K|$ being the volume
of cell $K$ and $\vect{w}_l^K =
(w_{\mathrm{\rho},l},\vect{w}_{\mathrm{p},l},w_{\mathrm{e},l})$ being the
$l$-th component of the weight vector of the density, momentum density and
total energy density. The integrals can be either solved analytically or
numerically using Gauss quadrature rules. By $w_{i,l}$ we refer to a
single component of the $l$-th weight vector, i.e.~$w_{0,0}$ and $w_{0,1}$
are the zeroth and first weights of the density field, which correspond to
the mean density and a quantity proportional to the gradient inside a
cell, respectively. If polynomial basis functions of degree k are used, a
numerical scheme with spatial order p=k+1 is achieved. However,
near discontinuities such as shock waves, the convergence order breaks
down to first order accuracy. A set of test problems demonstrating the
claimed convergence properties of our implementation can be found in
\cite{Schaal2015}.

\subsection{Basis functions}
We discretize the computational domain with a Cartesian grid and adopt a
classical modal DG scheme, in which the solution is given as a linear
combinations of orthonormal basis functions $\phi_l^K$. For the latter we
use tensor products of Legendre polynomials. The cell extensions are
rescaled such that they span the interval from $-1$ to $1$ in each
dimension. The transformation is given by \begin{align}
\vect{\xi}=\frac{2}{\Delta x^K}\left(\vect{x}-\vect{x}^K\right),
\end{align} with $\vect{x}^K$ being the centre of cell $K$.

The full set of basis functions can be written as
\begin{align}
\left\{\phi_l(\vect{\xi})\right\}_{l=1}^{N(k)}=\left\{\tilde{P}_u(\xi_1)\tilde{P}_v(\xi_2)\tilde{P}_w(\xi_3)|u,v,w\in\mathbb{N}_0\wedge u+v+w\le k\right\},
\end{align}
where $\tilde{P}_u$ are scaled Legendre polynomials of degree $u$. The
sum of the degrees of the individual basis functions has to be equal
or smaller than the degree $k$ of the DG scheme. Thus, the vector
space of all polynomials up to degree $k$ has the dimensionality
\begin{align}
N(k)=\sum\limits_{u=0}^k\sum\limits_{v=0}^{k-u}\sum\limits_{w=0}^{k-u-v}1=\frac{1}{6}(k+1)(k+2)(k+3).
\end{align}

\subsection{Initial conditions}

To obtain the initial conditions, we have to find weight vectors
$\vect{w}_l^K$ at $t=0$ corresponding to the initial conditions
$\vect{q}(\vect{x},0)$. The polynomial representation of a scalar
quantity described by the weight vector is
\begin{align}
q_{i}^K\left(\vect{x},0\right)=\sum\limits_{l=1}^{N(k)}w_{i,l}^K(0)\phi_l^K(\vect{x}).
\end{align}
The difference between the prescribed actual initial condition and the polynomial
representation should be minimal, which can be achieved by varying the
weight vectors $\vect{w}^K_l$ in each cell for each hydrodynamical
component $i$ individually:
\begin{align}
\min_{\left\{w^K_{i,l}(0)\right\}_l}\int_{K}\!\left(q_i^K(\vect{x},0)-q_i(\vect{x},0)\right)^2\,\mathrm{d}V,
\end{align}
Thus, the $l$-th component of the initial weights $\vect{w}_l^K$ is given
by
\begin{align}
\vect{w}^K_l(0)=\frac{1}{|K|}\int_K\!\vect{q}(\vect{x},0)\phi^K_l(\vect{x})\,\mathrm{d}V.
\end{align}
Transformed into the $\xi$ coordinate system, the equation becomes
\begin{align}
\vect{w}^K_l(0)=\frac{1}{8}\int_{[-1,1]^3}\!\vect{q}(\vect{\xi},0)\phi_l(\vect{\xi})\,\mathrm{d}\vect{\xi}.
\end{align}
In principle, the integral can be computed analytically for known
analytical initial conditions. Alternatively, it can be computed
numerically using a Gauss quadrature rule:
\begin{align}
\vect{w}^K_l(0)\cong\frac{1}{8}\sum_{q=1}^{(k+1)^3}\vect{q}(\vect{x}_q,0)\phi_l(\vect{\xi}_q)\omega_q,
\end{align}
using $(k+1)^3$ sampling points $\vect{x}_q$ and corresponding quadrature
weights $\omega_q$. 
With $k+1$ integration points polynomials of degree $\leq 2k+1$ are
integrated exactly by the Gauss quadrature rule. Therefore, the projection integral is exact for initial conditions
in the form of polynomials of degree $\leq k$.

\subsection{Time evolution equations} 

The solution is discretized using time-dependent weight vectors
$\vect{w}_l^K (t)$. The time evolution equations for these weights can
be derived from the Euler equation,
\begin{align}
\frac{\partial\vect{q}}{\partial{t}}+\sum\limits_{\alpha=1}^{3}\frac{\partial\vect{F}_{\alpha}(\vect{q})}{\partial x_{\alpha}}=0.
\end{align}
To obtain an evolution equation for the $l$-th weight, the Euler
equation is multiplied with $\phi_l$ and integrated
over the the volume of cell $K$,
\begin{align}
\frac{\mathrm{d}}{\mathrm{d}t}\int_K\!\vect{q}^K\phi_l^K\,\mathrm{d}V-\sum_{\alpha=1}^3\int_K\!\frac{\partial \vect{F}_{\alpha}(\vect{q})}{\partial x_\alpha} \phi_l^K\,\mathrm{d}V=0.
\end{align}
Integrating the second term by parts and applying Gauss's theorem leads to
a volume integral over the interior of the cell and a surface integral
with surface normal vector $\vect{n}$:
\begin{align}
\frac{\mathrm{d}}{\mathrm{d}t}\int_K\!\vect{q}^K\phi_l^K\,\mathrm{d}V-\sum_{\alpha=1}^3\int_K\!\vect{F}_\alpha\frac{\partial \phi_l^K}{\partial x_\alpha}\,\mathrm{d}V
+\sum_{\alpha=1}^3\int_{\partial K}\vect{F}_\alpha\phi_l^K n_{\alpha}\,\mathrm{d}A=0.
\end{align}

We will now discuss the three terms in turn, starting with the first
one.  Inserting the definition of $\vect{q}^K$ and using the
orthogonality relation of the basis functions simplifies this term to
the time derivative of the $l$-th weight:
\begin{align}
\frac{\mathrm{d}}{\mathrm{d}t}\int_K\!\vect{q}^K\phi_l^K\,\mathrm{d}V=|K|\frac{\mathrm{d}\vect{w}_l^K}{\mathrm{d}t}.
\label{dg:eqn:time_deriv}
\end{align}

We transform the next term into the $\xi$-coordinate system. The term
involves a volume integral, which is solved using a Gauss quadrature rule:
\begin{align}
\nonumber&\sum_{\alpha=1}^3\int_K\!\vect{F}_\alpha \left (\vect{q}^K \left(\vect{x},t\right) \right)\frac{\partial \phi_l^K(\vect{x})}{\partial x_\alpha}\,\mathrm{d}V\\
\nonumber=&\frac{\left(\Delta x^K\right)^2}{4}\sum_{\alpha=1}^3\int_{[-1,1]^3}\!\vect{F}_\alpha \left(\vect{q}^K \left(\vect{\xi},t \right)\right)\frac{\partial \phi_l(\vect{\xi})}{\partial \xi_\alpha}\,\mathrm{d}\vect{\xi}\\
\cong&\frac{\left(\Delta x^K\right)^2}{4}\sum_{\alpha=1}^3\sum_{q=1}^{(k+1)^3}\vect{F}_\alpha \left(\vect{q}^K \left(\vect{\xi}_q,t \right) \right) \left.\frac{\partial \phi_l}{\partial \xi_\alpha} \right|_{\vect{\xi}_q}\omega_q.
\label{dg:eqn:volume_int}
\end{align}
The flux vector $\vect{F}_{\alpha}$ can be easily evaluated at the
$(k+1)^3$ quadrature points $\vect{\xi}_q$ using the polynomial
representation $\vect{q}^K(\vect{\xi}_q,t)$. An analytical expression can
be obtained for the derivatives of the basis functions.

Finally, the last term is a surface integral over the cell
boundary. Again, we transform the equation into the $\xi$-coordinate
system and apply a Gauss quadrature rule to compute the integral:
\begin{align}
\nonumber &\sum_{\alpha=1}^3\int_{\partial K}\vect{F}_\alpha\phi_l^K(\vect{x})n_{\alpha}\,\mathrm{d}A\\
\nonumber = &\frac{\left(\Delta x^K\right)^2}{4}\int_{\partial [-1,1]^3}\numflux \left(\vect{q}_L^K(\vect{\xi},t),\vect{q}_R^K(\vect{\xi},t)\right)\phi_l(\vect{\xi})n_{\alpha}\,\mathrm{d}A'\\
\cong&\frac{\left(\Delta x^K\right)^2}{4}\sum_{a \in \partial [-1,1]^3}\sum_{q=1}^{(k+1)^2}\numflux \left(\vect{q}_L^K(\vect{\xi}_{a,q},t),\vect{q}_R^K(\vect{\xi}_{a,q},t)\right)\phi_l(\vect{\xi}_{q})\omega_{a,q}.
\label{dg:eqn:surface_int}
\end{align}
Each of the interface elements $a$ is sampled using $(k+1)^2$ quadrature
points $\vect{\xi}_{a,q}$. The numerical flux $\numflux$ between the
discontinuous states at both sides of the interface $\vect{q}_L^K$ and $\vect{q}_R^K$ is computed using an
exact or approximative HLLC Riemann solver. Note that only this term
couples the individual cells with each other.

Equations~(\ref{dg:eqn:volume_int}) and (\ref{dg:eqn:surface_int}) can be
combined into a function $\vect{R}_l^K \left
(\vect{w}_1,\ldots,\vect{w}_{N(k)}\right)$. Combining this with
Eq.~(\ref{dg:eqn:time_deriv}) gives the following system of coupled
ordinary differential equations for the weight vectors
$\vect{w}_l^K$:
\begin{align}
\frac{\mathrm{d}\vect{w}_l^K}{\mathrm{d}t}+\vect{R}_l^K\left(\vect{w}_1,\dots,\vect{w}_{N(k)}\right)=0.
\label{dg:eqn:time}
\end{align}
We integrate Eq.~(\ref{dg:eqn:time}) with an explicit strong
stability preserving (SSP) Runge-Kutta scheme \cite{Gottlieb2001}. We
define $\tensor{y} = \left (\vect{w}_1,\dots,\vect{w}_{N(k)} \right)$ and
thus we have to solve
\begin{align}
\frac{\mathrm{d} \tensor{y}}{\mathrm{d}t}+R(\tensor{y})=0.
\end{align}
A third order SSP Runge-Kutta scheme used in our implementation is given by
\begin{align}
\tensor{y}^{(0)} &= \tensor{y}^n \\
\tensor{y}^{(1)} &= \tensor{y}^{(0)} - \Delta t^n R(\tensor{y}^{(0)}) \\
\tensor{y}^{(2)} &= \frac{3}{4} \tensor{y}^{(0)} + \frac{1}{4} \left( \tensor{y}^{(1)} - \Delta t^n R(\tensor{y}^{(1)}) \right) \\
\tensor{y}^{(3)} &= \frac{1}{3} \tensor{y}^{(0)} + \frac{2}{3} \left( \tensor{y}^{(2)} - \Delta t^n R(\tensor{y}^{(2)}) \right) \\
\tensor{y}^{n+1} &= \tensor{y}^{(3)}.
\end{align}
with initial value $\tensor{y}^n$, final value $\tensor{y}^{n+1}$,
intermediate states $\tensor{y}^{(0)}, \tensor{y}^{(1)},\tensor{y}^{(2)}$,
and time step size $\Delta t^n$.

\subsection{Time-step calculation}
The time step has to fulfill the following Courant criterium
\cite{Cockburn1989a}:
\begin{align}
\Delta t^K=\frac{\textrm{C}}{2k+1}\left(\frac{|v_1^K|+c^K}{\Delta x_1^K}+\frac{|v_2^K|+c^K}{\Delta x_2^K}+\frac{|v_3^K|+c^K}{\Delta x_3^K}\right)^{-1},
\end{align} 
with Courant factor C, components of the mean velocity $v_i^K$ in
cell $K$ and sound speed $c^K$. The minimum over all cells is determined
and taken as the global maximum allowed time step. Note the $(2k +
1)^{-1}$ dependence of the time step, which leads to a reduction of the
timestep for high order schemes.

\subsection{Positivity limiter} Higher order methods usually
  require some form of limiting to remain stable. However there is no
  universal solution to this problem and the optimum choice of such a
  limiter is in general problem dependent. For our set of turbulence
simulations we have decided to limit the solution as little as
possible and adopt only a positivity limiter. This choice may lead to
some oscillations in the solution, however, it achieves the most
accurate result in terms of error measurements. We explicitly
  verified this for the case of shock tube test problems. At all
times, the density $\rho$, pressure $P$ and energy $e$ should remain
positive throughout the entire computational domain. However, the
higher order polynomial approximation could violate this physical
constraint in some parts of the solution. This in turn can produce a
numerical stability problem for the DG solver if the positivity is
violated at a quadrature point inside the cell or an interface. To
avoid this problem, we use a so-called positivity limiter
\cite{Zhang2010}. By applying this limiter at the beginning of
  each Runge-Kutta stage, it is guaranteed that the density and
  pressure values entering the flux calculation are positive, as well
  as the mean cell values at the end of each RK stage. In addition, a
  strong stability preserving Runge-Kutta scheme and a positivity
  preserving Riemann solver is needed to guarantee positivity.

The set of points where positivity is enforced has to include the cell
interfaces, because fluxes are computed there as well. A possible choice of
integration points, which include the integration edges, are the
Gauss-Lobatto-Legendre (GLL) points. In the following, we will be using
tensorial products of GLL and Gauss points, where one coordinate is chosen
from the set of GLL points and the remaining two are taken from the set of
Gauss points:
\begin{align}
S_x=\{(\hat{\xi}_r, \xi_s, \xi_t): 1\le r \le m, 1\le s \le k+1, 1\le t \le k+1\}\\
\nonumber S_y=\{(\xi_r, \hat{\xi}_s, \xi_t): 1\le r \le k+1, 1\le s \le m, 1\le t \le k+1\}\\
S_z=\{(\xi_r, \xi_s, \hat{\xi}_t): 1\le r \le k+1, 1\le s \le k+1, 1\le t \le m\}
\end{align}
The full set of integration points is $S = S_x \cup S_y \cup S_z$, which
includes all points where fluxes are evaluated in the integration step.

First, the minimum density at all points in the set $S$ is computed:
\begin{align}
\rho_{\text{min}}^K=\min_{\vect{\xi}\in S} \rho^K(\vect{\xi}).
\end{align}
We define a reduction factor $\theta_1^K$ as
\begin{align}
\theta_1^K=\min\left\{\left|\frac{\bar{\rho}^K-\epsilon}{\bar{\rho}^K-\rho_{\text{min}}^K}\right|,1\right\}, 
\end{align}
with the mean density in the cell $\bar{\rho}^K$ (the 0-th density weight)
and the minimum target density $\epsilon$. All high order weights of the
density are reduced by this factor
\begin{align}
w_{j,1}^K\leftarrow\theta_1^K w_{j,1}^K,\quad j=2,...,N(k).
\end{align}
To guarantee a positive pressure $P$, a similar approach is taken:
\begin{align}
\theta_2^K=\min_{\vect{\xi}\in S} \tau^K(\vect{\xi}),
\end{align}
with
\begin{align}
\tau^K(\vect{\xi})=
\begin{cases}
1\quad&\text{if}\,\,\,P^K(\vect{\xi})\ge \epsilon \\
\tau_\ast\quad&\text{such that}\,\,\,P(\vect{q}^K(\vect{\xi})+\tau_\ast(\vect{q}^K(\vect{\xi})-\bar{\vect{q}}^K))=\epsilon.
\end{cases}
\label{eq:p_positivity}
\end{align}
The equation for $\tau$ can not be solved analytically and has to be solved
numerically. To this end we employ a Newton-Raphson method. Now,
the higher order weights of all quantities are reduced by $\theta_2$
\begin{align}
w_{j,i}^K\leftarrow\theta_2^K w_{j,i}^K,\quad j=2,...,N(k),\quad i=1,...,5.
\end{align}
Additionally the timestep has to be modified slightly to
\begin{align}
\Delta t^K=\text{C}\, \min\left(\frac{1}{2k+1}, \frac{\hat{w}_1}{2}\right)
\left(\frac{|v_1^K|+c^K}{\Delta x_1^K}+\frac{|v_2^K|+c^K}{\Delta x_2^K}+\frac{|v_3^K|+c^K}{\Delta x_3^K}\right)^{-1},
\label{eq:time_step_pos}
\end{align}
with the first GLL weight $\hat{w}_1$. For a second order DG scheme the
first weight is $\hat{w}_1 = 1$, and $\hat{w}_1 = 1/3$ for a third and
fourth order method.

\section{Turbulence simulations}

We shall consider an effectively isothermal gas in which we drive
subsonic turbulence through an external force field on large scales. The imposed
isothermality prevents the buildup of internal energy and pressure
through the turbulent cascade over time. Technically, we simulate an
ideal gas but reset slight deviations from isothermality back to to
the imposed temperature level after every timestep, allowing us to
directly measure the dissipated internal energy. 

We consider a 3D simulation domain of size $L = 1$.  In the
following, we will compare runs with a finite volume scheme and runs
using our new DG hydro solver on a fixed Cartesian mesh. In the case
of DG simulations we vary the resolution as well as the convergence
order of the code. A summary of all of our runs is given in
Table~\ref{turb:tab:summary}. 

Note that we always state the convergence order,
i.e.~$\mathcal{O} = k+1$ instead of $k$ for our DG runs. At a fixed
convergence order of $3$, we vary the resolution from $32^3$ up to
$256^3$, and at a fixed resolution of $128^3$ we change the
convergence order from $1$ up to $4$. This allows us to asses the
impact of both parameters against each other. The number of basis
functions is $N(0) = 1$ for a first order method, $N(1)=4$ for a
second order method, $N(2)=10$ for a third order, and $N(3) = 20$ for
a fourth order method. In Table~\ref{turb:tab:summary} we also state
the approximate number of degrees of freedom per dimension to better
compare the impact of increasing the order versus increasing the
resolution level. We compare against a second order  MUSCL type finite volume method, using an exact Riemann solver.

\begin{table*}
  \begin{center}
\begin{tabular}{lcccc}
\hline
\multicolumn{4}{l}{Overview over our turbulence simulations}\\ 
\hline
Label & Numerical method & Conv.\ order $\mathcal{O}$ & Resolution &  $(\mathrm{d.o.f.}/\mathrm{cell})^{1/3}$\\
\hline
\newlength{\myl}\settowidth{\myl}{DG}\parbox{\myl}{FV}\_{\em X}\_1  & finite volume & $1$ & $32^3$ \dots $512^3$ & $1$\\
\newlength{\myll}\settowidth{\myll}{DG}\parbox{\myll}{FV}\_{\em X}\_2  & finite volume & $2$ & $32^3$ \dots $512^3$ & $1$\\
DG\_{\em X}\_1  & discontinuous Galerkin & $1$ & $128^3$ & $1$\\
DG\_{\em X}\_2  & discontinuous Galerkin & $2$ & $128^3$ & $1.59$\\
DG\_{\em X}\_3  & discontinuous Galerkin & $3$ & $32^3 \dots 256^3$ & $2.15$\\
DG\_{\em X}\_4  & discontinuous Galerkin & $4$ & $128^3$ & $2.71$\\
\hline
\end{tabular}
\caption[Summary of the turbulence simulations]{Summary of the turbulence
simulations discussed in this article. The {\em X} in the name is a
placeholder for the resolution level. As a reference solution we consider
ordinary finite volume simulations with up to $512^3$ resolution elements.
In case of DG, we vary the resolution from $32^3$ up to $256^3$ for the
third order code, as well as the convergence order from $1$ up to $4$ at a
resolution of $128^3$ cells. To better asses the impact of a higher order
method, we state the number of degrees of freedom per cell per dimension.
The number of degrees of freedom per cell are $1, 4, 10$ and $20$ (from
$1$ order up to $4$ order) in the case of DG}
\label{turb:tab:summary}
\end{center}
\end{table*}

\subsection{Turbulence driving}

We use the same driving method as in \cite{Bauer2012}, which is based
on \cite{Schmidt2006,Federrath2008a,Federrath2009,Federrath2010} and
\cite{Price2010}. We generate a turbulent acceleration field in
Fourier space containing power in a small range of modes between
$k_{\mathrm{min}}=6.27$ and $k_{\mathrm{max}}=12.57$. The amplitude of
the modes is described by a paraboloid centered around
$(k_{\mathrm{min}}+k_{\mathrm{max}})/2$.  The phases are drawn from an
Ornstein--Uhlenbeck (OU) process. This random process is given by
\begin{align}
\vect{\theta}_{t}=f\, \vect{\theta}_{t-\Delta t} + \sigma \sqrt{(1-f^2)}\, \vect{z}_n ,
\end{align}
with random variable $\vect{z}_n$ and decay factor $f$, given by
$f=\exp(-\Delta t/t_s)$, with correlation length $t_s$. The phases are
updated after a time interval of $\Delta t$. The variance of the process
is set by $\sigma$. The expected mean value of the sequence is zero,
$\left <\vect{\theta}_t\right >=0$, and the correlations between random
numbers over time are $\left<\vect{\theta}_t\,\vect{\theta}_{t+\Delta
t}\right> = \sigma^2 f$. This guarantees a smooth, but frequent change of
the turbulent driving field.

We want a purely solenoidal driving field, because we are interested in
smooth subsonic turbulence in this study. A compressive part would only
excite sound waves, which would eventually steepen to shocks if the
driving is strong enough. These compressive modes are filtered out through
a Helmholtz decomposition in Fourier space:
\begin{equation}
  \vect{\hat a}(\vect{k})_i =\left(\delta_{ij} - \frac{k_i k_j}{|{k}|^2}\right) \vect{\hat a_0}(\vect{k})_j.
\end{equation}

The acceleration field is incorporated as an external source term in
the DG equations. The formalism is similar to adding an external
gravitational field. We need to compute the following DG integrals for
$\vect{a}_l^K$:
\begin{align}
\nonumber\vect{a}_l^K(t) = &\int_K\!\vect{a}(\vect{x},t)\phi_l^K({\vect{x}})\,\mathrm{d}V\\
\nonumber=&\frac{|K|}{8}\int_{[-1,1]^3}\!\vect{a}(\vect{\xi},t)\phi_l(\vect{\xi})\,\mathrm{d}\vect{\xi}\\
\cong&\frac{|K|}{8}\sum_{q=1}^{(k+1)^3}\!\vect{a}(\vect{\xi}_q,t)\phi_l(\vect{\xi}_q)\omega_q,
\end{align}
thus we have to evaluate the driving field for $(k+1)^3$ inner quadrature
points $\vect{\xi}$ for each Runge-Kutta stage. An additional evaluation
at the cell centre is required to compute the allowed time step size. A corresponding term is used to update the energy equation as well. The
evaluation is done with a discrete Fourier sum over the few non-zero modes
of the driving field. If the update frequency of the driving field is
smaller than the typical timestep size, storing the acceleration field for
each inner quadrature point can speed up the computations. In case of the
finite volume runs, we add the driving field through two half step kick
operators at the beginning and end of a time step, like for ordinary
gravity.

The overall amplitude of the acceleration field is rescaled such that a
given Mach number is reached. Our target Mach number is $\mathcal{M} \sim
0.2$. The decay time scale is chosen as half the eddy turnover time scale,
$t_s = \frac{1}{2} \frac{L}{{\mathcal{M}c}} = 2.5$ in our case. The
acceleration field is updated $10$ times per decay time scale, $\Delta t =
0.1 t_s = 0.25$.

\begin{figure}
    \sidecaption[t]
\includegraphics[width=7.5cm]{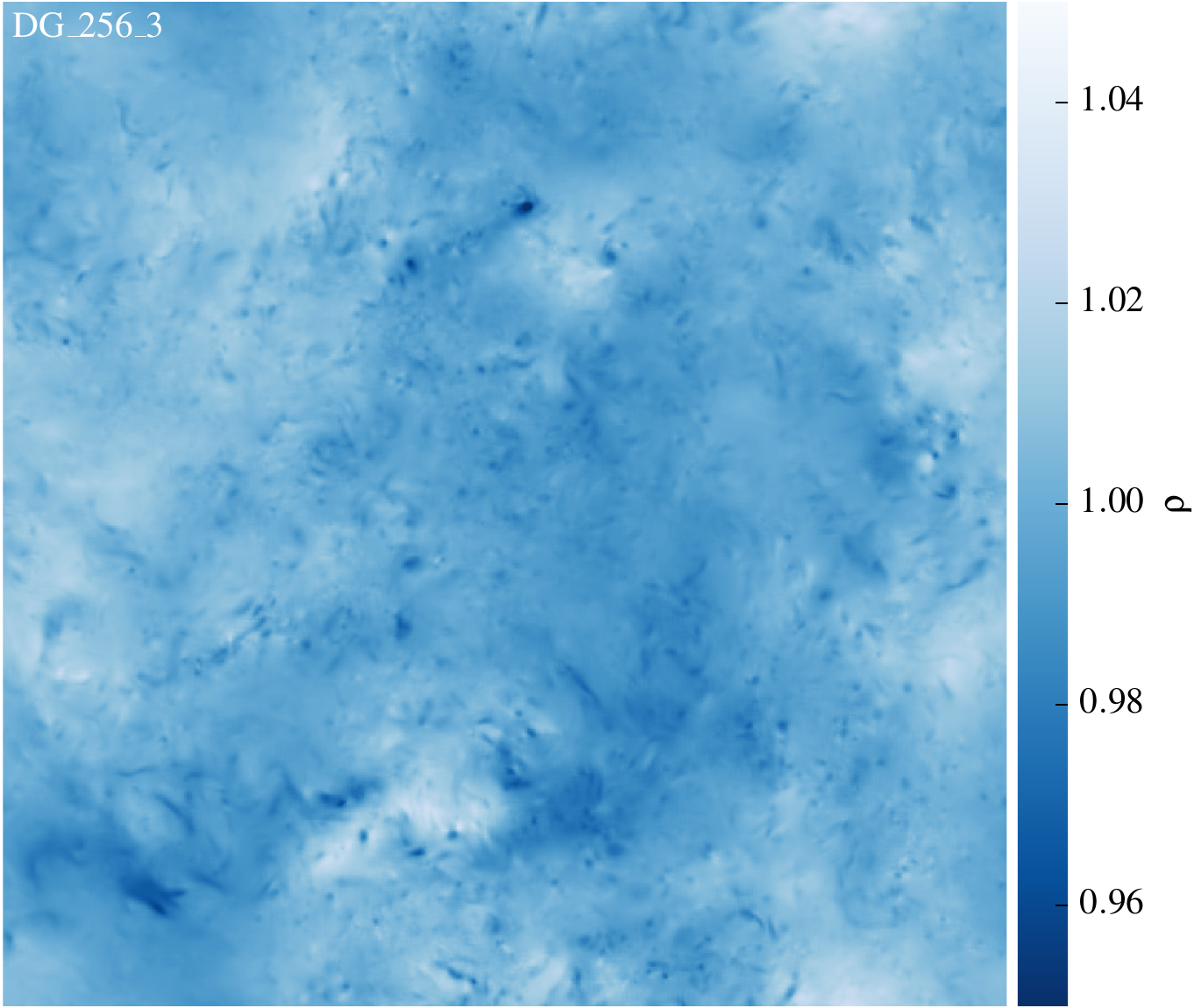}%
\caption{A thin slice through the middle of our best resolved DG
simulation at third order showing the density field. The field uses the
sub-cell information given by the high order DG weights. Every cell is
sub-sampled four times}
\label{turb:fig:slice_large}
\end{figure}

\subsection{Dissipation measurement}

We use an adiabatic index of $\gamma = 1.01$ instead of the isothermal
index $\gamma = 1$. The slight deviation from $\gamma = 1$ allows us to
measure the dissipated energy while the dynamics of the fluid is
essentially isothermal. After each timestep, the expected specific
internal energy is computed as
\begin{align}
  \epsilon = \frac{c^2}{\gamma -1} \frac{\rho^{\gamma -1}}{\rho_0^{\gamma - 1}},
\end{align}
with sound speed $c$ and reference density $\rho_0 = 1$. This specific
internal energy is enforced at all quadrature points within a cell. Thus,
the weights associated with the total energy density using the kinetic
momentum and density field have to be adjusted:
\begin{align}
\nonumber w_{\mathrm{e},l}^K(t) = &\int_K\!\left ( \frac{1}{2} \frac{\vect{p}(\vect{x},t)^2}{\rho(\vect{x},t)} + \rho(\vect{x},t) \epsilon(\vect{x},t) \right) \phi_l^K({\vect{x}})\,\mathrm{d}V\\
\nonumber=&\frac{|K|}{8}\int_{[-1,1]^3}\!\left ( \frac{1}{2} \frac{\vect{p}(\vect{\vect{\xi}},t)^2}{\rho(\vect{\xi},t)} + \rho(\vect{\xi},t) \epsilon(\vect{\xi},t) \right)\phi_l(\vect{\xi})\,\mathrm{d}\vect{\xi}\\
\cong&\frac{|K|}{8}\sum_{q=1}^{(k+1)^3}\!\left ( \frac{1}{2} \frac{\vect{p}(\vect{\vect{\xi}_q},t)^2}{\rho(\vect{\xi}_q,t)} + \rho(\vect{\xi}_q,t) \epsilon(\vect{\xi}_q,t) \right)\phi_l(\vect{\xi}_q)\omega_q.
\end{align}
Afterwards, the average internal energy density in the cell can be
recomputed as
\begin{align}
  \rho \epsilon = {w_{\mathrm{e},0}^K} - \frac{1}{2} \frac{{\vect{w}_{\mathrm{p},0}^K}^2}{w_{\mathrm{\rho},0}^K}.
\end{align}
The dissipated energy is given by the difference between the average
internal energy before and after adjusting the weights of the total energy
density. Afterwards the positivity limiter is applied to guarantee
non-negative values in our DG simulations.

\subsection{Power spectrum measurement}
\begin{figure}
\begin{center}
\includegraphics[width=12cm]{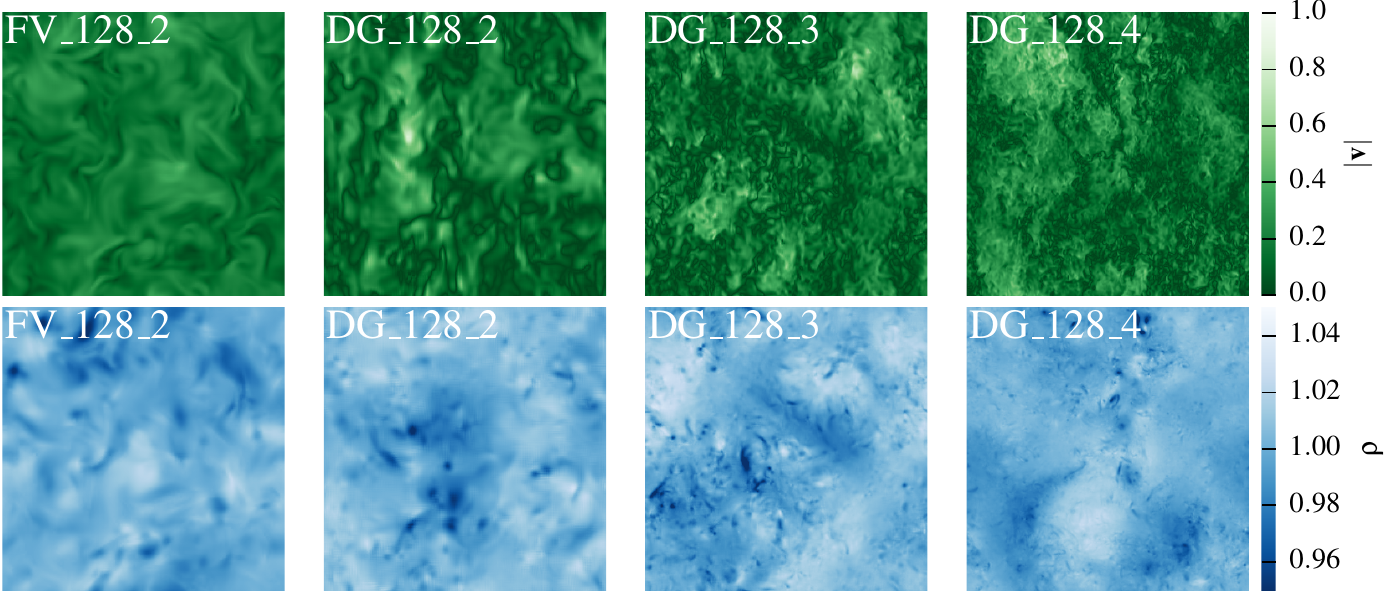}%
\caption{Thin slices through the density and velocity field at $t=30$. We
compare the finite volume simulations against DG simulations of order 2 up
to 4. Already 2nd order DG shows features which are finer than in the 2nd
order finite volume run. The higher moments available in 3rd and 4th order
DG allow a representation of finer features without increasing the spatial
resolution. The thin lines of zero velocity are much more pronounced in
case of DG than in the finite volume case
\label{turb:fig:slice}}
\end{center}
\end{figure}

The power spectrum of a scalar or vector field $w(\vect{x})$ is
proportional to the Fourier transformed of the two point correlation
function:
\begin{align}
C_w(\vect{l}) = \langle  w(\vect{x}+\vect{l})w(\vect{x}) \rangle_{\vect{x}}.
\end{align}
Thus
\begin{align}
E_w(\vect{k}) &= (2 \pi)^{3/2} \mathcal{F}(C_w(\vect{l})) = \int_V C_w(\vect{l}) \exp (-i
\vect{k} \vect{l})\, {\mathrm{d}}^3\vect{l}\\
&= \left |\hat{w}({\vect k}) \right |^2 ,
\end{align}
where $\hat{w}$ is the Fourier transform of $w$\footnote{We are using
  the convention of normalizing the Fourier transform symmetrically
  with $(2 \pi)^{-3/2}$.}.  Here, we are only interested in the 1D
power spectrum, thus we average $E_w({\vect k})$ over spherical
shells:
\begin{align}
E_w(k) = 4 \pi k^2 \langle E_w({\vect k}) \rangle ,
\end{align}
where $k = \left | {\vect k} \right |$. The overall normalization of the
Fourier transformation is chosen such that the integral over the power
spectrum is equivalent to the total energy:
\begin{align}
  \sigma^2  = \int w(x) \, {\mathrm{d}}\vect{x} = \int E_w(k) \, {\mathrm{d}}k = \frac{1}{(2\pi)^3 N^3}
  \sum_{i,j,k=0}^{N-1} |\hat{w}_{ijk}|^2,
\end{align}
with $\hat{w}_{ijk}$ being the discrete Fourier transformation of the
discretized continuous field $w$.  Usually we show $kE(k)$ instead of
$E(k)$ directly in log-log plots. This means a horizontal line in a
log-log plot represents equal energy per decade and makes interpreting
the area under a curve easier.

\section{Results}

In Fig.~\ref{turb:fig:slice} we show a first visual overview of our
simulation results at a resolution of $128^3$ cells. The panels show
the state at the final output time $t=30$ for the magnitude of the
velocity and the density in a thin slice through the middle of the
box. Each cell is subsampled four times for this plot using the full
sub-cell information present for each DG or finite volume cell. In the
case of the finite volume scheme, we used the estimated gradients in
sub-sampling the cells.

The finite volume and DG results are similar at second order accuracy.
However, already the second order DG run visually shows more small scale
structure than the finite volume run. By increasing the order of accuracy
and therefore allowing for more degrees of freedom within a cell, DG is
able to represent considerably more structure at the same number of cells.
Interestingly, the velocity field has regions of (almost) zero velocity.
These thin stripes can be well represented in DG. The finite volume run
shows the same features, but they are not as pronounced.
Additionally, Fig.~\ref{turb:fig:slice_large} shows a thin density slice
for our highest resolution DG run DG\_256\_3. The high resolution and
third order accuracy allows for more small scale details than in any other
of our simulations.

\subsection{Mach number evolution}

\begin{figure}
    \sidecaption[t]
\includegraphics[width=7.5cm]{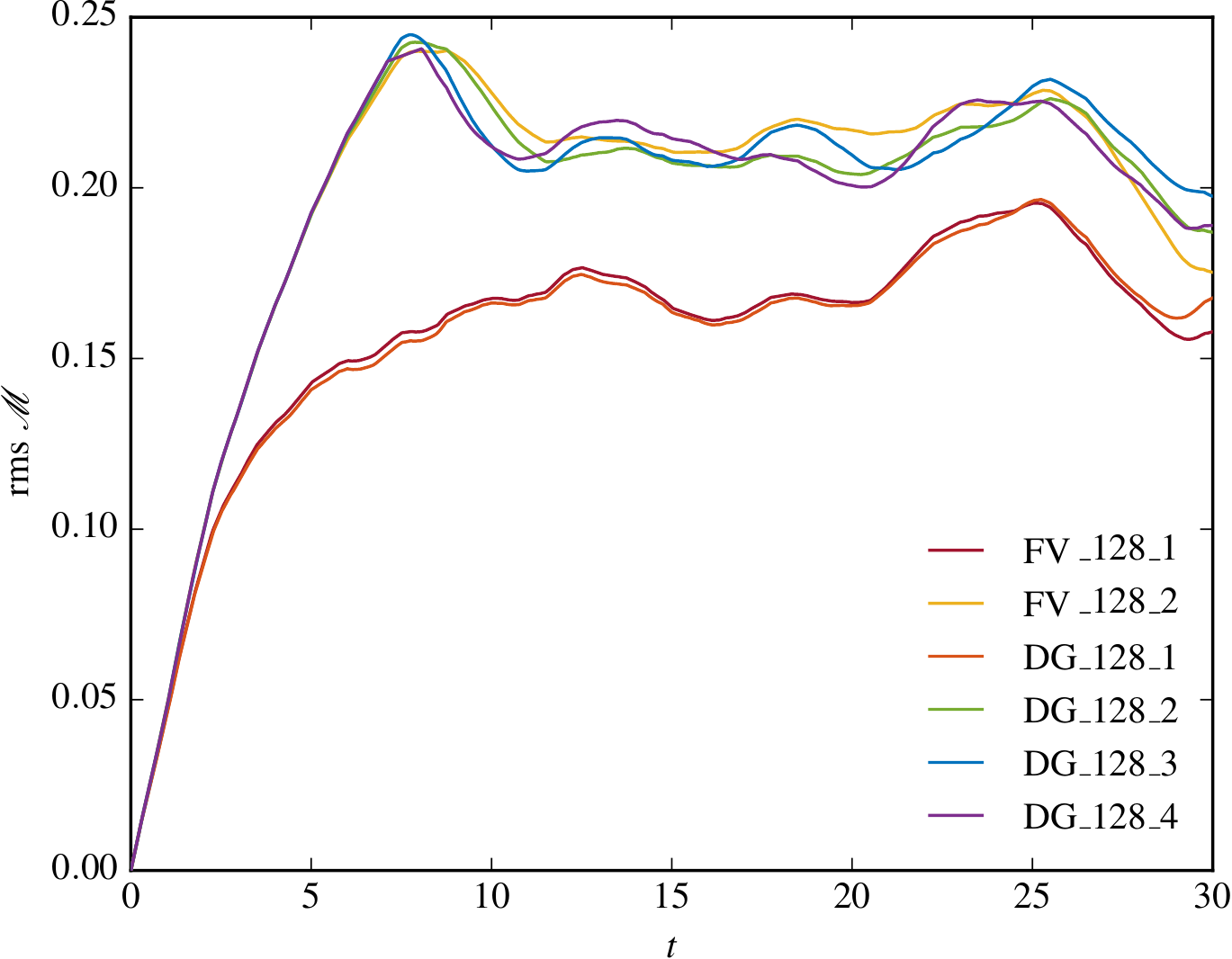}%
\caption{Time evolution of the root mean square Mach number
  $\mathcal{M}$.  The runs with a higher than first order convergence
  order agree well with each other and establish a Mach number of
  about $\mathcal{M} \sim 0.2$ at $t=12$ in the quasi-stationary
  phase. However, the first order finite volume and DG runs do not
  manage to reach the same Mach number and fall substantially short of
  achieving a comparable kinetic energy throughout the entire run
  time}
\label{turb:fig:mach}
\end{figure}

All of our runs with a convergence order larger or equal to second
order reach an average Mach number of $\mathcal{M} \sim 0.21$ after
$t=12$. The detailed history of the Mach number varies a bit from run
to run. The differences between the different DG runs and finite
volume runs are however insignificant. The same holds true for the
other runs not shown in Fig.~\ref{turb:fig:mach}. Interestingly,
both, the first order finite volume and DG runs fall substantially
behind and can only reach a steady state Mach number of about
$\mathcal{M} \sim 0.17$. The low numerical accuracy leads to a too
high numerical dissipation rate in this case, preventing a fully
established turbulent cascade. A similar problem was found in
\cite{Bauer2012} for simulations with standard SPH even at
comparatively high resolution, caused by a high numerical viscosity
the noisy character of SPH.

\subsection{Injected and dissipated energy}
\begin{figure}
    \sidecaption[t]
\includegraphics[width=7.5cm]{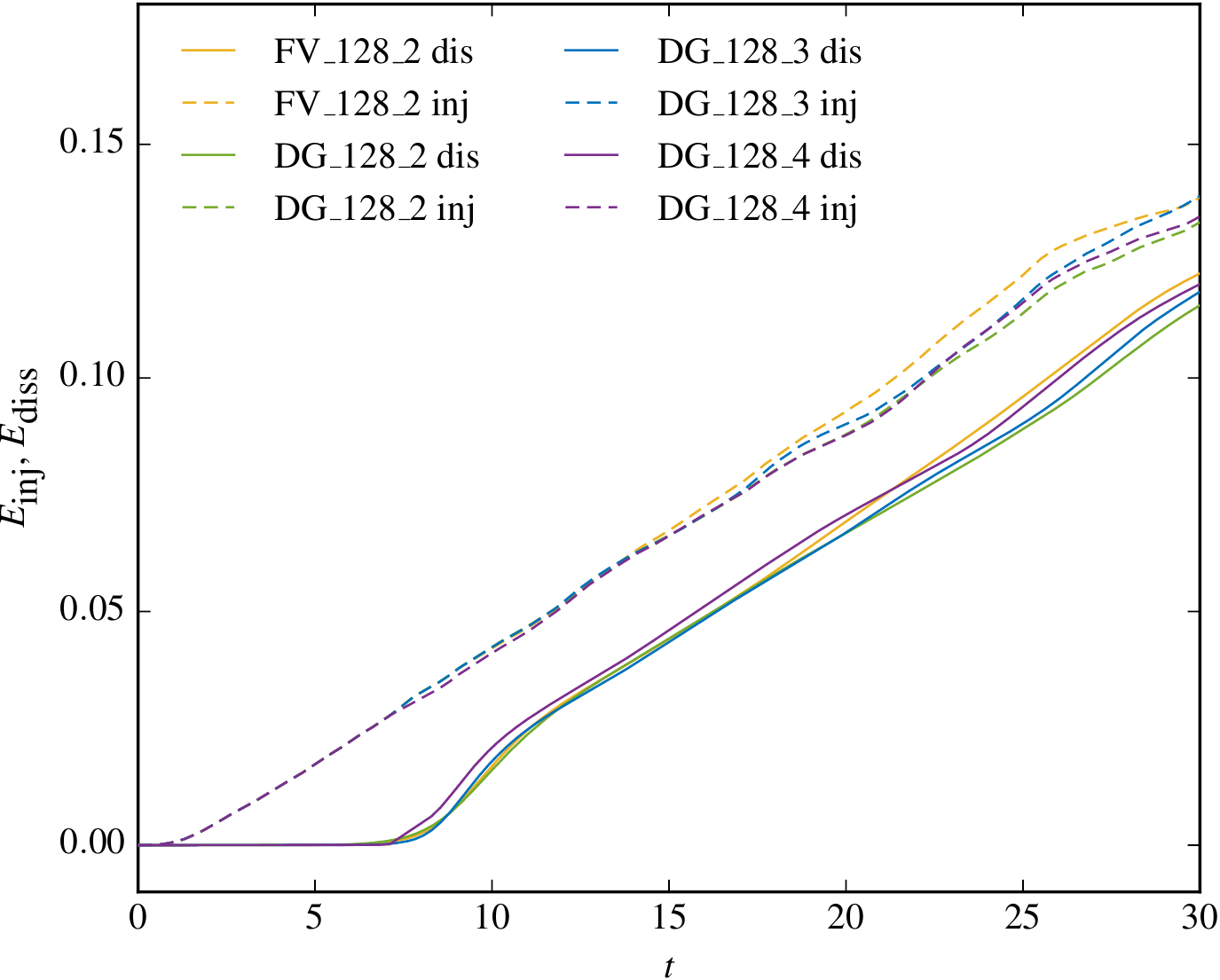}%
\caption{The dashed lines show the injected energy, while the solid
  lines give the dissipated energy over time. Dissipation becomes only
  relevant after an initial start-up phase. Thereafter, a
  quasi-stationary state is established}
\label{turb:fig:inj}
\end{figure}

The globally injected and dissipated energy in our turbulence
simulations is shown in Fig.~\ref{turb:fig:inj} as a function of
time. The rate of energy injection through the driving forces stays
almost constant over time. At around $t=12$, the variations start to
increase slightly. At this point the fluctuations between individual
runs start to grow as well. Initially, the dissipation is negligible,
but at around $t=8$ dissipation suddenly kicks in at a high rate, and
then quickly transitions to a lower level at around $t=12$, where a
quasi stationary state is reached that persists until the end of our
runs. The difference between both curves -- the kinetic energy --
remains rather constant after $t=12$. Thus, in the following we only
use outputs after $t=12$ for our analysis.

\subsection{Velocity power spectra}
\label{turb:sec:power}

\begin{figure}
    \sidecaption[t]
\includegraphics[width=7.5cm]{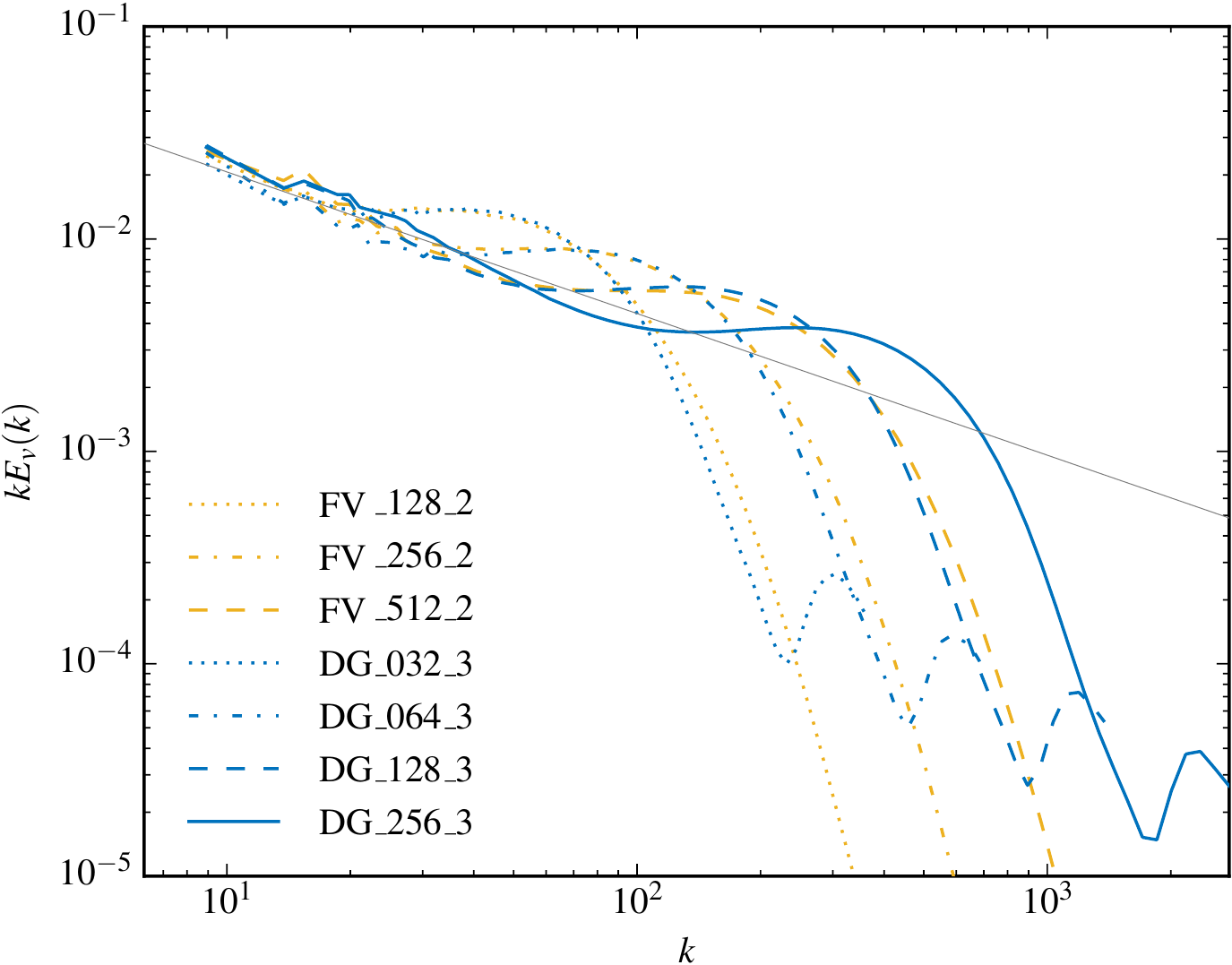}%
\caption{Comparison of the velocity power spectra of our second order
  finite volume runs against our third order DG runs. Interestingly,
  the spectra of the DG runs match with the ones obtained from the
  finite volume runs at a quarter of the resolution. Thus, DG obtains
  similar results using only about half as many degrees of freedom per
  dimension as finite volume schemes. For comparison, the grey line shows the $k^{-5/3}$ Kolmogorov scaling}
\label{turb:fig:power_resolution}
\end{figure}

\begin{figure}
    \sidecaption[t]
\includegraphics[width=7.5cm]{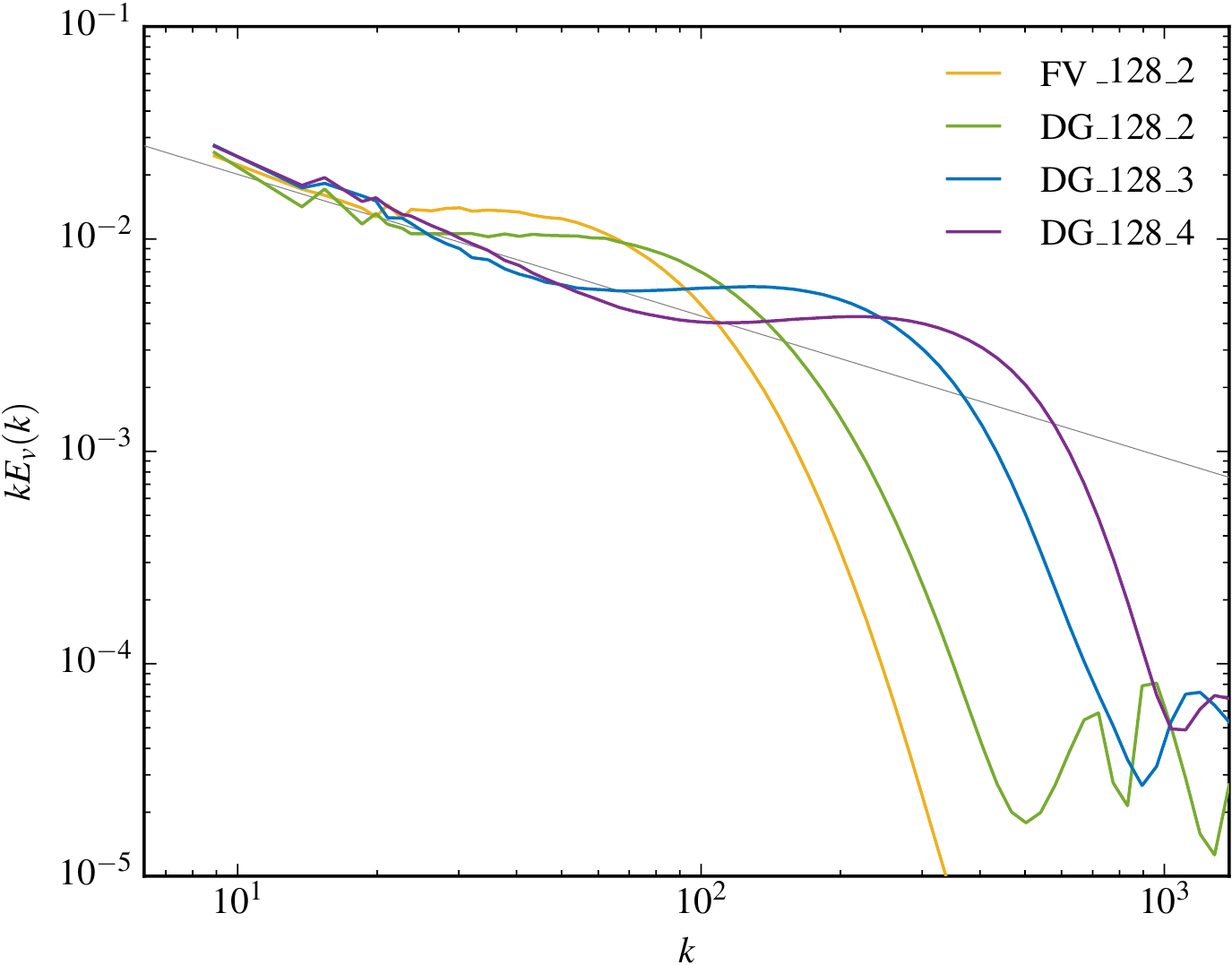}%
\caption{Velocity power spectrum for our DG runs at different
convergence order at a resolution of $128^3$ cells. Already second order
DG shows a large inertial range and a dissipation bottleneck at small
scales. For comparison, the grey line shows the $k^{-5/3}$ Kolmogorov scaling}
\label{turb:fig:power_order}
\end{figure}

In Fig.~\ref{turb:fig:power_resolution} and
\ref{turb:fig:power_order}, we show velocity power spectra of our
runs. First, we focus on a resolution study of our third order DG and
second order finite volume simulations in
Fig.~\ref{turb:fig:power_resolution}. In case of the finite volume
runs, we show the power spectra up to the Nyquist frequency
$k_n=2 \pi N/ 2 L$, with $N$ being the number of cells per
dimension. For our DG runs we show the full power spectrum instead,
obtained from the grid used in the Fourier transformation up to
$k_{g} = 2 \pi 4 N /2 L = 4 k_n$. The finite volume runs have a second
peak not shown here at modes above $k_n$, induced by noise resulting
from the discontinuities across cell boundaries. The third and higher
order DG methods show a still declining power spectrum at $k_n$ and
only at even higher modes close to $k_g$ start to show a noise induced
rise. This is due to the available sub-cell information encoded in the
DG weights.

All runs show an inertial range at scales smaller than the driving
range on large scales. The inertial range is followed by a numerical
dissipation bottleneck. This bottleneck is similar to the
experimentally observed physical bottleneck effect, but appears to be
somewhat stronger. The energy transfered to smaller scales can not be
dissipated fast enough at the resolution scale and piles up there
before it is eventually transformed to heat. The bottleneck feature
moves to ever smaller scales as the numerical resolution is
increased. Especially our highest resolution DG run $DG\_256\_3$ shows
a quite large inertial range. However, the slope of the inertial range
is measured slightly steeper than the expected $k^{-5/3}$ Kolmogorov
scaling. We think a Mach number of $\mathcal{M}\sim 0.21$ and the
associated density fluctuations are maybe already too high for a
purely Kolmogorov-like turbulence cascade, which is only expected for
incompressible gas.

Interestingly, the power spectra of our finite volume runs match those
of our third order DG simulations, except that the finite volume
scheme requires four times higher spatial resolution per
dimension. Considering the $10$ degrees of freedom per cell for third
order DG, the effective number of degrees of freedom is still lower by
a factor of $6.4$ in the case of DG, which corresponds to a factor of
$1.86$ per dimension. This underlines the power of higher order
numerical methods, especially if comparatively smooth problems such as
subsonic turbulence are studied.

In Fig.~\ref{turb:fig:power_order} we compare the impact of the numerical
convergence order on the power spectrum of our DG runs. As a comparison we
include a second order finite volume run as well. All simulations have a
numerical resolution of $128^3$. Already the second order DG method shows
a more extended inertial range than the second order finite volume
run. But the second order DG method already uses four degrees of
freedom per cell. Increasing the convergence order alone improves the
inertial range considerably. The change in going from second to third
order is a bit larger than the change from third to fourth order.

\subsection{Density PDFs}

\begin{figure}
    \sidecaption[t]
\includegraphics[width=7.5cm]{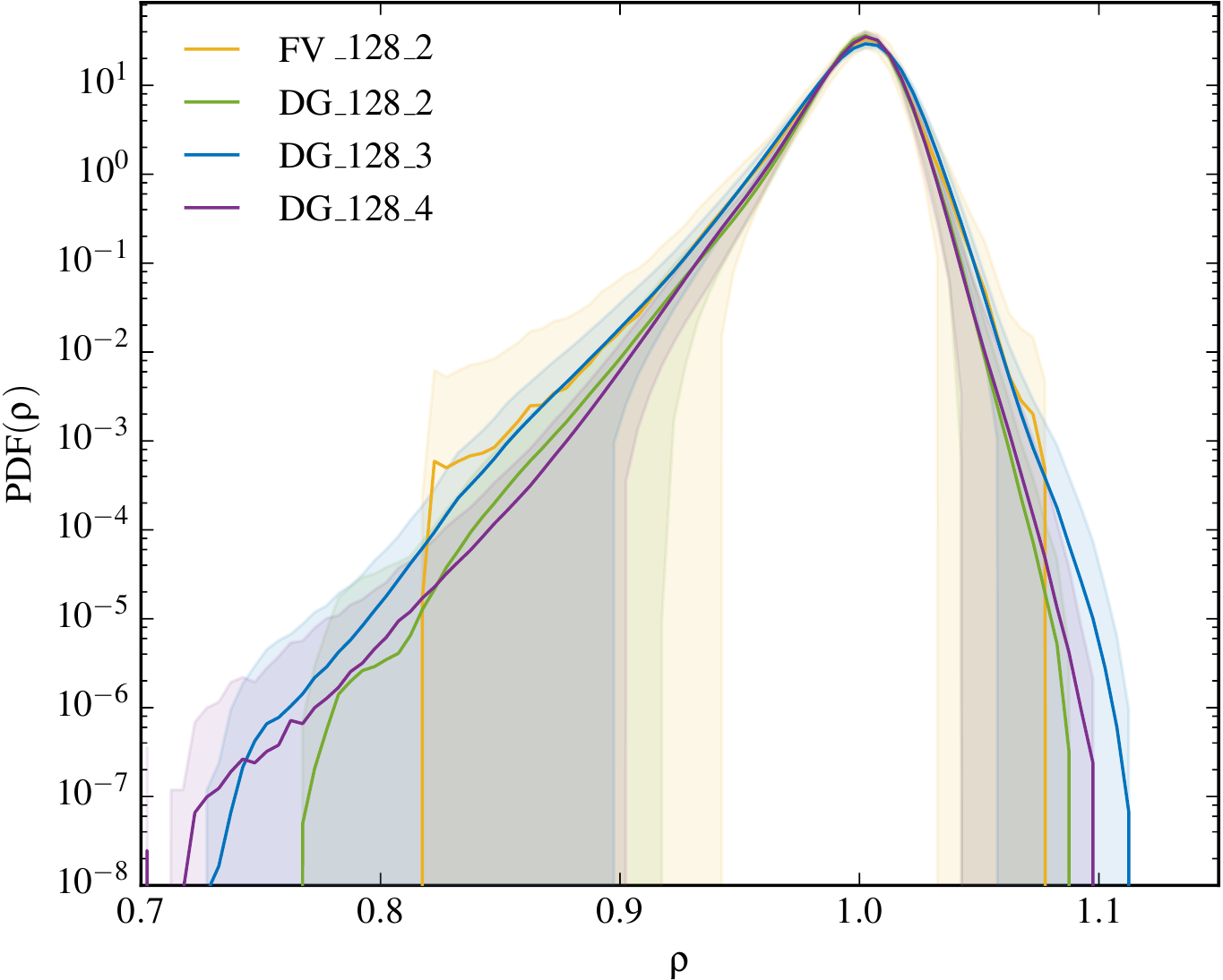}%
\caption{The density PDF for our runs at a resolution of $128^3$
  cells.  The PDF is obtained by subsampling each cell $4^3$ times. In
  the finite volume case, we take the estimated density gradients into
  account. For DG, we use the full polynomial information present in
  each cell. The shaded area represents the standard deviation over
  time. Interestingly, finite volume schemes show a sharp drop off at the low
  and high density ends which is absent in this form in the DG calculations}
\label{turb:fig:pdf}
\end{figure}

In Fig.~\ref{turb:fig:pdf}, we show the probability density function (PDF)
of the density field for some of our runs. The PDF is averaged from $t=12$
up to $t=30$ and sub-sampled $4^3$ times for each cell. We take the
estimated density gradients into account for the finite volume runs. The
finite volume run shows the smallest range of realized density values at
the sampling points. Slightly more sampling points pile up at the extreme
density values. This is due to the slope limited gradients used here,
preventing more extreme density values. The DG runs show a more extended
range of density values, with the range increasing with convergence order,
because the higher order polynomial representations allow for a more
detailed structure with more extrema within a cell. If only the mean
values within the cells are considered, the PDFs are all rather similar to
each other and not so different from the finite volume run shown.

\section{Discussion}

We presented the ideas and equations behind a new implementation of
discontinues Galerkin hydrodynamics that we realized in the
astrophysical simulation code {\small TENET}
\cite{Schaal2015}. Unlike traditional finite volume schemes, DG uses
subcell expansion into a set of basis functions, which leads to
internal flux calculations in addition to surface integrals that are
solved by a Riemann solver. Importantly, the reconstruction step
needed in finite volume schemes is obsolete in DG. Instead, the
coefficients of the expansion are evolved independently, and no
information is `thrown away' at the end of a timestep, unlike done by
the implicit averaging in finite volume schemes at the end of every
step. This offers the prospect of a higher computational efficiency,
especially at higher order where correspondingly more information is
retained from step to step. Such higher order can be relatively easily
achieved in DG approaches. Furthermore, stencils only involve direct
neighbours in DG, even for higher order, thereby making parallelization
on distributed memory systems comparatively easy and efficient.

These advantages clearly make DG methods an interesting approach for
discretizing the Euler equations. However, at shocks, the standard
method may fall back to first order accuracy unless sophisticated
limiters are used. If the problem at hand is dominated by shocks and
discontinuities, this might be a drawback. Ultimately, only detailed
application tests, like we have carried out here, can decide which
method proves better in practice.

In this regard, an important question for comparing numerical methods
is their computational efficiency for a given accuracy, or conversely,
what is the best numerical accuracy which can be obtained for a given
invested total runtime. Obtaining a fair comparison based on the run
time of a code can be complicated in general. For example, for the
runs analyzed in this study, different numbers of CPUs had to be used,
as the memory requirements change by several orders of magnitude
between our smallest and largest runs. The comparison may be further
influenced by the fact that both hydro solver implementations
investigated here are optimized to different degrees (with much more
tuning already done for the finite volume method), which can distort
simple comparisons of the run times. Nevertheless, we opted to give a
straightforward comparison of the total CPU time used as a first rough
indicator of the efficiency of our DG method compared to a
corresponding finite volume scheme. We note however that our new DG
code is less optimized thus far compared with the finite volume
module, so we expect that there is certainly room for further
improving the performance ratio in favor of DG.

We performed our simulations running in parallel on up to 4096
  cores on {\small SUPERMUC}. In part thanks to the homogenous
  Cartesian mesh used in these calculations, our code showed excellent
  strong parallel scaling. We note that this is far harder to achieve
  when the adaptive mesh refinement (AMR) option present in {\small
    TENET} is activated.

We have generally found that the DG results for subsonic turbulence
are as good as the finite volume ones, but only need slightly more
than half as many degrees of freedom for comparable accuracy.  Both
the finite volume method at second order accuracy and the DG scheme at
third order accuracy show very good weak scaling when increasing the
resolution for the range of resolutions studied here. If we compare
the run time for roughly equal turbulence power spectra, we find that
the DG\_032\_3 run is about $1.14$ times faster than the corresponding
FV\_128\_2 run. This performance ratio increases if we improve the
resolutions: The DG\_064\_3 is already $1.34$ times faster than the
FV\_256\_2 run. The DG\_128\_3 run is $1.53$ times faster than the
FV\_512\_2 run, which comes close to the factor $1.86$ more degrees of
freedom needed in the finite volume run to achieve the same
accuracy. Thus, DG does not only need fewer degrees of freedom to
obtain the same accuracy but also considerably less run time. This
combination makes DG a very interesting method for solving the Euler
equations.

Besides improving computational efficiency, DG has even more to
offer. In particular, it can manifestly conserve angular momentum in
smooth parts of the flow, unlike traditional finite volume methods. In
addition, the ${\rm div} \vec{B}= 0$ constraint of ideal
magnetohydrodynamics (MHD) can be enforced at the level of the basis
function expansion, opening up new possibilities to robustly implement
MHD \cite{Mocz2014}.  Combined with its computational speed, this
reinforces the high potential of DG as an attractive approach for
future exascale application codes in astrophysics, potentially
replacing the traditional finite volume scheme that are still in
widespread use today.

\begin{acknowledgement}
  We thank Gero Schn\"ucke, Juan-Pablo Gallego, Johannes L\"obbert,
  Federico Marinacci, Christoph Pfrommer, and Christian Arnold for
  very helpful discussions. The authors gratefully acknowledge the
  support of the Klaus Tschira Foundation.  We acknowledge financial
  support through subproject EXAMAG of the Priority Programme 1648
  `SPPEXA' of the German Science Foundation, and through the European
  Research Council through ERC-StG grant EXAGAL-308037.  KS and AB
  acknowledge support by the IMPRS for Astronomy and Cosmic Physics at
  the Heidelberg University. PC was supported by the AIRBUS 
Group Corporate Foundation Chair in Mathematics of Complex 
Systems established in TIFR/ICTS, Bangalore.
\end{acknowledgement}

\bibliography{turb}
\bibliographystyle{spphys}
\end{document}